\documentclass[fleqn,usenatbib]{mnras}

\usepackage[T1]{fontenc}

\DeclareRobustCommand{\VAN}[3]{#2}
\let\VANthebibliography\thebibliography
\def\thebibliography{\DeclareRobustCommand{\VAN}[3]{##3}\VANthebibliography}

\usepackage{graphicx}	
\usepackage{amsmath}	
\usepackage{amssymb}	
\usepackage{newtxtext,newtxmath}
\usepackage[usenames,dvipsnames]{xcolor}
\usepackage{orcidlink}
\usepackage{balance}

\hypersetup{
    colorlinks = true,
    citecolor = {MidnightBlue},
    linkcolor = {BrickRed},
    urlcolor = {BrickRed}
}

\title[Beam Inference from Visibilities]{High Dimensional Beam Inference II: Inference of a Perturbed HERA Beam from Simulated Visibility Data}

\author[M. J. Wilensky et al.]{
Michael J. Wilensky,\orcidlink{https://orcid.org/0000-0001-7716-9312}$^{1,2, \dag}$\thanks{E-mail: \url{michael.wilensky@mcgill.ca}}
Philip Bull,\orcidlink{0000-0001-5668-3101}$^{1,3}$
Nicolas Fagnoni${^4}$\,\orcidlink{0000-0001-5300-3166}
\\
$^{1}$Jodrell Bank Centre for Astrophysics, University of Manchester, Manchester M13 9PL, UK\\ 
$^{2}$Department of Physics and Trottier Space Institute, McGill University, 3600 University Street, Montreal, QC H3A 2T8, Canada\\
$^{3}$Department of Physics and Astronomy, University of Western Cape, Cape Town 7535, South Africa\\
${^4}$Cavendish Astrophysics, University of Cambridge, Cambridge CB3 0HE, UK\\
$^{\dag}$ CITA National Fellow
}

\pubyear{2025}

\begin{document}
\label{firstpage}
\pagerange{\pageref{firstpage}--\pageref{lastpage}}
\maketitle

\begin{abstract}
Accurate beam modeling is important in many radio astronomy applications. In this paper, we focus on beam modeling for 21-cm intensity mapping experiments using radio interferometers, though the techniques also apply to single dish experiments with small modifications. In 21-cm intensity mapping, beam models are usually determined from highly detailed electromagnetic simulations of the receiver system. However, these simulations are expensive, and therefore have limited ability to describe practical imperfections in the beam pattern. We present a fully analytic Bayesian inference framework to infer a beam pattern from the interferometric visibilities assuming a particular sky model and that the beam pattern for all elements is identical, allowing one to capture deviations from the ideal beam for relatively low computational cost. We represent the beam using a sparse Fourier-Bessel basis on a projection of the hemisphere to the unit disc, but the framework applies to any linear basis expansion of the primary beam. We test the framework on simulated visibilities from an unpolarized sky, ignoring mutual coupling of array elements. We successfully recover the simulated, perturbed power beam when the sky model is perfect. Briefly exploring sky model inaccuracies, we find that beam inferences are sensitive to them, so we suggest jointly modeling uncertainties in the sky and beam in related inference tasks.
\end{abstract}

\begin{keywords}
methods: statistical -- methods: data analysis
\end{keywords}



\section{Introduction}

Radio astronomers must be able to characterize the electromagnetic responses of their receiving elements i.e. their \textit{beams}. This can be done in a number of ways, depending on the dynamic range of the particular observational campaign, available resources, the observing wavelength, etc. The dynamic range is a particularly strong driver of the characterization process, since higher precision characterization demands more effort. For example, when making maps of Galactic diffuse emission using single dish surveys, it is important to convert the map to the appropriate scale using an accurate value for the solid angle of the Stokes I ``power beam'' \citep{Jonas1998, Remazeilles2015}. Inaccurate characterization of the primary beam in this instance can result in an error in the flux scale of the map. However, sub-percent level precision in this scale, which may require detailed knowledge of the far sidelobes, may be unnecessary for e.g. characterizing the brightness Galactic synchrotron emission \citep{Kogut2012, Irfan2022, Wilensky2025}. In high dynamic range (HDR) cosmology experiments, such as in the search for cosmic HI emission, improperly characterized sidelobes can completely ruin the measurement \citep{Ewall-Wice2017, Matshawule2021, Barry2022, Chokshi2024, Cumner2024}. In this work, we focus on such HDR applications, with particular emphasis on use of large radio arrays.

It is common to use detailed electromagnetic simulations to characterize the beam pattern of a given receiving element \citep{Asad2021, Fagnoni2021a, Fagnoni2021b, Cumner2022, Cumner2024, Sampath2024, RHINO}. If these simulations are sufficiently detailed, then they can be used directly in analysis pipelines without biasing the measurement. However, knowing whether the model is sufficiently accurate requires one of two things, which are not mutually exclusive concepts. Either the beam needs to be measured \textit{in situ} to the dynamic range required, or beam needs to be \textit{inferred} as a free parameter during astrophysical/cosmological inference. We focus on the latter option in this paper. 

Beam measurement has been explored in terms of holographic measurements \citep{VLA2019, Asad2021, CHIME2022}, measurements based on satellite passage \citep{Line2018, Chokshi2024}, and drone-based measurements \citep{2014IAWPL..13..169V, 2015ExA....39..405P, echo, Tyndall2024}. All of the aforementioned cases of measurement explicitly rely on the calibration of the partner antenna (holography, satellite passage), or the transmitter, and are thus limited by this in some cases. On the other hand, the radio sky is generally stable night-to-night, implying that one can (theoretically) integrate very deeply to obtain extremely sensitive characterizations of the beam through an \textit{in situ} method such as this one. However one is still relying on an accurate model of the transmitting source i.e. the sky. If the analysis is sufficiently advanced, then in principle this can even be done with significant errors in the \textit{a priori} sky model as long as the sky model can be updated jointly \textit{a posteriori} along with the beam \citep{Kern2025}, though we do not want to understate the difficulty of this task. 

Bayesian inference is a formal way of quantifying statistical uncertainty of model parameter values inferred from noisy measurements \citep{Jaynes, Gelman2021}. The advantage of fully Bayesian modeling, compared to e.g. forming point estimates and using standard but \textit{ad hoc} error propagation rules that are valid only in a limited context, is that the uncertainty is accurately propagated through the complicated, often nonlinear relationships between different model components in the form of a joint probability distribution. This allows the different properties of the data to then be reported in a statistically unified way that reflects these complicated relationships.  Uncertain prior knowledge (e.g. of source catalogs) can also be included in a self-consistent way. In high dynamic range problems such as 21cm analysis, small inaccuracies and neglected uncertainties can significantly bias the recovered signal; a fully Bayesian approach guards against such biases by accurately incorporating the model uncertainty into the signal estimates. The radio sky and telescope beam patterns are generally complicated, potentially requiring large numbers of parameters \citep[e.g. hundreds to thousands depending on desired accuracy;][]{Wilensky2024, Sampath2024, Glasscock}. Constraining such parametrized models in a fully Bayesian framework can be computationally challenging for this reason, and methods are adopted to minimize the dimensionality of these models through approximate inference \citep{Cumner2024, BGSM}.  Nevertheless, more exact high-dimensional models may be necessary for high dynamic range applications like interferometric measurements of the 21-cm power spectrum, especially in the presence of systematic effects like mutual coupling \citep{Bolli2023, Rath2025} that can produce per-antenna effects in the instrument response. It is therefore important to explore high-dimensional inference techniques in general. 

Necessary steps towards this goal include inferring a sky, given a beam and some measurements (i.e. making a map), which is more or less synonymous with the practice of radio astronomy and has been widely explored in numerous contexts. Particular advances that are highly relevant to our approach and important for e.g. 21-cm cosmology involve low-frequency Galactic diffuse modeling \citep{Glasscock, Eastwood2018, Byrne2022b, Yatawatta2024}, and in particular statistically based map-making strategies like Direct Optimal Mapping \citep{Dillon2015, Xu2022}, m-mode analysis \citep{Shaw2015, Eastwood2018}, and Bayesian approaches \citep{Lochner2015, Roth2023, Glasscock}. 

Complementary to the mapping step is the beam inference step, i.e. inferring a beam from e.g. visibility measurements and a given sky model. This is not an entirely novel concept \citep{Maaskant2012, Yatawatta2012, Young2013, Cumner2024}, and is essentially \textit{direction-dependent calibration} \citep[e.g.][]{Yatawatta2018, Yatawatta2021, Gan2023, Roth2023, Brackenhoff2025}. What we introduce in this work is a Bayesian technique that works for very high-dimensional beam models (thousands of parameters) based on linear basis expansions. The necessity for individually validated beam models in the search for the cosmic 21-cm signal during the Epoch of Reionization was explored by \citet{Chokshi2024}, and generally beam patterns within an interferometric array may vary element-to-element\footnote{By \textit{element} we specifically mean the components whose signals are sent to the correlator. For example, each of HERA's elements involve one dish and one crossed dipole feed, while each of the MWA's elements in this context are themselves a phased array of dipoles in a square 4x4 configuration, often called a tile. It would be the entire tile's beam that would be described by this formalism, rather than a single member of the phased array.} due to effects such as environmental factors and limited engineering precision \citep{Orosz_2019, Choudhuri+21, Kim2022, Kim2023}. While we acknowledge the importance of per-element beam inference, the foundation of visibility based beam constraints is yet to be deeply explored, and so we use this work to set some of that foundation under the assumption of identical receiving elements.

Describing beam patterns with linear basis expansions has been shown to provide percent level approximations to simulated beams with tens to hundreds of coefficients \citep{bui2017,Asad2021, Sampath2024}, and with even sparser models in some SVD-based techniques \citep{Cumner2024}. This work is a continuation of \citep{Wilensky2024}, henceforth Paper I, where a sparse linear basis was found for the Hydrogen Epoch of Reionization Array \citep[HERA;][]{deBoer2017, Berkhout2024} single-element beam. Paper I involved modeling the fully-polarized Jones matrix of a HERA receiving element. In order for a simpler showcase of this new Bayesian technique, we focus on a toy model using the basis from Paper I and ignore polarization effects. This allows us to explore the properties of the inference in the absence of large complications associated with the more realistic case. The general methods in this paper for beam inference are not particular to the choice of basis i.e. we present a formalism where the chosen linear basis is arbitrary.

Since the posterior is fully analytic conditional on other observational parameters (e.g. direction independent calibration, sky model), this technique can be folded into a Gibbs sampling framework \citep{Gibbs}, and is itself part of a larger 21-cm Gibbs sampling framework called \textsc{Hydra}.\footnote{\url{https://github.com/HydraRadio/Hydra}} This head of \textsc{Hydra} is named \textsc{HydraBeam}. In general, Gibbs sampling is most effective when the conditional distributions of a complicated joint distribution are themselves tractable, such as when they are multivariate Gaussian. In this special case, high-dimensional parameter spaces can be sampled relatively quickly. Though, like many Markov Chain Monte Carlo (MCMC) methods, it is still fundamentally a random walk through the parameter space and so its exploration in terms of number of samples can still be slow \citep{Mackay2003}. However, if each individual sample can be made quickly due to conveniences in the form of the conditional probability distributions, then exploration of enormous parameter spaces can still be rapid in terms of total compute time. This paper essentially demonstrates and explores one step of \textsc{Hydra}, namely sampling from the beam model conditional on a sky model and other instrumental parameters like direction-independent gains. We reserve the incorporation of multiple heads of \textsc{Hydra} into a Gibbs sampler for future work, but offer a primer on Gibbs sampling in Appendix \ref{app:Gibbs} to establish future context for the interested reader. The main point is that the beam inference framework here is modular in the sense that it can be easily combined with a sky inference framework such as by \citet{Glasscock} to enable joint inference. However, even in the absence of joint modeling of these other observational parameters, the inference framework we present here will still be a useful tool for understanding the relationship between beams and visibility data. 

We explain our mathematical formalism in \S\ref{sec:math}, including the nature of the beam model and the statistical analysis of the simulated visibility data. We describe the simulation properties in \S\ref{sec:sims}, followed by a mock data analysis in \S\ref{sec:results}, including an investigation of the effect of an incomplete sky model. We conclude in \S\ref{sec:conc}.

\section{Mathematical Formalism}
\label{sec:math}

Here we discuss the physical assumptions, mathematical models, and statistical framework for analyzing inferred beams from visibility data.

\subsection{Visibility Model}


In general, the visibility for antennas $j$ and $k$ at time, $t$, frequency $\nu$, are modeled as,
\begin{multline}
        \boldsymbol{\mathcal{V}}_{jk}(\nu, t) = \int \mathrm{d}\Omega~ \boldsymbol{\mathcal{J}}_{j}(\hat{s}, \nu)\boldsymbol{\mathcal{C}}(\hat{s}, \nu, t)\boldsymbol{\mathcal{J}}^\dag_{k}(\hat{s}, \nu)e^{2\pi i \hat{s}\cdot\Delta\vec{x}_{jk} \nu / c},
        \label{eq:full_jones_vis}
\end{multline}
where $\boldsymbol{\mathcal{J}}_{j}(\hat{s}, \nu)$ is a direction-dependent Jones matrix that projects from some polarization basis on the sky to the instrumental polarization \citep{Jones1941}, $\hat{s}$ is a unit vector pointing to source positions on the sky in a zenith-centered frame, and $\mathrm{d}\Omega$ refers to the integration measure of a sphere. The term $\boldsymbol{\mathcal{C}}(\hat{s}, \nu, t)$ refers to the sky coherency, and $\exp(2\pi i \hat{s}\cdot\Delta\vec{x}_{jk} \nu / c)$ is the baseline fringe term, where $\Delta\vec{x}_{jk}$ is the difference in antenna positions. Here, $\boldsymbol{\mathcal{V}}_{jk}(\nu, t)$ is a $2\times2$ matrix representing the co-polarized visibilities on the diagonals and cross-polarized visibilities on the off-diagonals. The sky coherency matrix can be written in terms of Stokes polarization parameters, $(I, Q, U, V)$, and for an unpolarized sky, it is proportional to the identity matrix. If we assume antennas have identical beams, then a given instrumental polarization component, $\mathcal{V}{pqjk}(\nu, t) $, of Equation \ref{eq:full_jones_vis} can be written\footnote{The leading factor of $\frac{1}{2}$ is often treated as a conventional choice when normalizing visibilities during calibration. It arises from treating the sky coherency matrix as a covariance matrix of the electric field polarization components in a given basis, and then defining Stokes $I$ as the sum of the variances of the two polarization components. It might be omitted when, for example, only one instrumental polarization is used to make a Stokes $I$ image.}
\begin{equation}
    \mathcal{V}_{pqjk}(\nu, t) = \frac{1}{2}\int \mathrm{d}\Omega~ I(\hat{s}, \nu, t) B_{pq}(\hat{s}, \nu)e^{2\pi i \hat{s}\cdot\Delta\vec{x}_{jk} \nu / c},
    \label{eq:power_beam}
\end{equation}
where $B_{pq}(\hat{s}, \nu)$ is the $pq$ component of $\boldsymbol{\mathcal{J}}(\hat{s}, \nu)\boldsymbol{\mathcal{J}}^\dag(\hat{s}, \nu)$. This object is sometimes referred to as a ``power beam,'' particularly when $p=q$. Interestingly, this implies that the cross-polarized visibilities will be nonzero even for a pure Stokes $I$ sky, which further implies that information about the polarized response of the instrument can be inferred from the cross-polarized visibilities even without a polarized source to observe. This can be made sense of by noting that the projected instrumental polarization relative to a source can be non-orthogonal even if the dipole arms are physically orthogonal, which is most significant at the horizon \citep{Byrne2022}. We also note that with an unpolarized sky the Jones information is preserved only up to a unitary ambiguity since $\boldsymbol{\mathcal{J}}(\hat{s}, \nu)\boldsymbol{\mathcal{J}}^\dag(\hat{s}, \nu) = \boldsymbol{\mathcal{J}}(\hat{s}, \nu)\boldsymbol{\mathcal{U}}\boldsymbol{\mathcal{U}}^\dag\boldsymbol{\mathcal{J}}^\dag(\hat{s}, \nu)$ for any unitary matrix $\boldsymbol{\mathcal{U}}$ \citep[e.g.][]{Yatawatta2012, Yatawatta2018}. We only examine one polarization in \S\ref{sec:sims}-\ref{sec:results}, but remark on this here for the sake of future work.

From Equation \ref{eq:power_beam}, we can see that these assumptions establish a linear mapping between the power beam and the visibilities, conditional on a particular sky and antenna positions. While in theory sufficient measurements would allow one to constrain a ``per-pixel" beam by essentially imaging it, tighter constraints can be gained by acknowledging prior knowledge from electromagnetic simulations that beams are typically smooth functions, and can be compactly represented with linear superpositions of smooth basis functions \citep{Asad2021, Cumner2024, Sampath2024, Wilensky2024}. Mathematically, assuming
\begin{equation}
    B_{pq}(\hat{s}, \nu) = \sum_n b_{npq} f_n(\hat{s}, \nu)
    \label{eq:beam_expansion}
\end{equation}
for some set of basis functions, $f_n(\hat{s}, \nu)$, one may write Equation \ref{eq:power_beam} as
\begin{equation}
    \mathcal{V}_{pqjk}(\nu, t) = \frac{1}{2}\int \mathrm{d}\Omega~ I(\hat{s}, \nu, t) e^{2\pi i \hat{s}\cdot\Delta\vec{x}_{jk} \nu / c} \sum_n b_{npq} f_n(\hat{s}, \nu).
    \label{eq:lin_exp_vis}
\end{equation}
Strictly speaking, only a finite sum of basis functions can be used in application and it is not guaranteed that the physical beam is exactly represented by a finite sum. This means this relationship should really be regarded as approximate, and a robust modeling pipeline should examine the inference for signs of modeling error. We mock examination of that using our simulated data in \S\ref{sec:results}.

\subsection{Choice of Basis}

We use the Fourier-Bessel (FB) basis proposed in Paper I. This basis was chosen because it was able to accurately describe simulated beams for the HERA Phase I and Phase II receiving elements \citep{Fagnoni2021a, Fagnoni2021b} with many fewer basis functions than $1^\circ$ pixels on the sky (the simulation resolution). We refer the reader to Paper I for details on how a particular subset of FB modes are chosen. 

\newcommand{\bb}{\mathbfss{b}}

In paper I, we analyzed the Jones elements i.e. the fully polarized beam response. Here, we analyze power beams. We have seen that this basis also offers a compact representation of the simulated HERA power beams (both Phase I and Phase II). Each basis function is given by the product of a 0th order cylindrical Bessel function and a sinusoid:
\begin{equation}
    f_n(\theta, \phi) = \frac{J_0\left(u_{k_n}\rho\left(\theta\right)\right)e^{im_n\phi}}{q_{k_n}}
\end{equation}
where $\theta$ is the zenith angle coordinate, $\phi$ is the azimuth coordinate, the notation $k_n$ and $m_n$ indicates that there are unique pairs $(k, m)$  associated with each basis function, $u_{k_n}$ is the $k$th zero of the $J_0(x)$, $q_{k_n}$ is a normalization factor chosen for easier basis function comparison (cf. Paper I), and
\begin{equation}
    \rho(\theta) \equiv \frac{\sqrt{1 - \cos\theta}}{\alpha}
\end{equation}
for some constant $\alpha$ close to 1 (slightly tuned for better horizon behavior, see Paper I). Specifically we use
\begin{equation}
    \alpha = \sqrt{1 - \cos(23\pi/45)}.
\end{equation}
We use 1024 total basis functions.\footnote{There are roughly $20,000$ square degrees on a hemisphere so this still represents a factor of 20 spatial compression compared to a degree-scale pixel based representation.} 

Note we do not use frequency-dependent basis functions in this work. We only explore simulations at one frequency (150 MHz), but compression along the frequency axis using e.g. Legendre polynomials \citep{Kern2025} can be folded naturally into this formalism. Alternatively, the entire problem parallelizes over the frequency axis if no frequency-coupling is induced by the prior on $\bb$. This may be highly advantageous from a computational perspective, but HDR experiments with tight spectral demands such as the search for the cosmic 21cm signal may risk strong degeneracies between the target signal and the beam with the latter approach.

\newcommand{\FF}{\mathbfss{F}}

\subsection{Statistical Framework}

To simplify the analysis, we write the sky model as a sum of point sources. While in principle this formulation could simultaneously represent extragalactic point sources along with a pixelated form of Galactic diffuse emission, computational gains can be made by acknowledging that Galactic emission may be more compactly represented with a superposition of smooth functions \citep{Glasscock}. We only simulate a fictitious point source catalog in \S\ref{sec:sims}. Assuming circular Gaussian noise on each baseline, this formulation allows us to write a Gaussian linear model for the data, $\mathbfss{d}$:
\begin{equation}
    \mathbfss{d} | \mathbfss{b}, \mathbfss{N}, I \sim \mathcal{N}\left(\mathbfss{F}(I)\mathbfss{b}, \mathbfss{N}\right)
    \label{eq:likelihood}
\end{equation}
where $\mathbfss{N}$ is the noise covariance, and $\FF(I)$ indicates the design matrix is a function of the sky model. Componentwise,
\begin{equation}
    F_{jkpq\nu t n} = \frac{1}{2}\sum_m I_m(\hat{s}_m, \nu) e^{2\pi i \hat{s}_m\cdot\Delta\vec{x}_{jk} \nu / c} f_n(\hat{s}_m, \nu),
    \label{eq:forward_model}
\end{equation}
where $m$ indexes point sources in the sky model. There are several important notes about this expression. 

First, we have assumed the basis functions are independent of the polarization pair in consideration, $pq$, and an unpolarized sky. That means this design matrix, $\mathbfss{F}$, is constant over $p$ and $q$, and does not need to be recomputed every time. Since we only examine one polarization in this work, this is somewhat irrelevant here, but may be an important consideration for future work.

Second, the computation of the components is expressed as a sum over point sources, implicitly taking advantage of a Dirac-$\delta$ function representation. However, a more primitive form in terms of Equation \ref{eq:lin_exp_vis} shows that the construction of this design matrix only necessitates that a visibility simulation is performed once per beam basis function, where the ``beam'' in a given simulation is equal to the basis function. In other words, different implementations of visibility simulations may be exchanged at this step. This may be particularly advantageous to consider if faster visibility simulation implementations are available for a given sky model by taking advantage of nonuniform fast Fourier transforms \citep{FFTVIS}, parallelization \citep{Sullivan2012, pyuvsim, Woden, PRISM, CASA, matvis}, etc. This is the most computationally intense step in our analysis. For the small simulations in this work this only takes seconds in modern high performance computing environments with a relatively naive \textsc{numpy}-based implementation. However, folding this into a larger sampling framework such as \textsc{Hydra} may demand more optimal implementations of this step, since the design matrix (or one like it) may need to be computed for each posterior sample (cf. Appendix \ref{app:Gibbs}).

Finally, along with assuming a perfect sky model, we have assumed perfect direction-independent calibration. Generally a sky and beam model are used to generate calibration solutions, and so by itself this framework can be potentially biased by a previous calibration process that makes assumptions about the instrument beam. In a full joint modeling framework, which is the ultimate goal of the research arc of which this paper is only a small part, the joint posterior of the sky model, beam model, and direction-independent gains would be constrained by the data. The relationship between these model components would be summarized by degeneracies (correlations or more complicated statistical dependence in the joint posterior), resulting in larger uncertainties, but not necessarily \textit{biases} that typically result from strict assumptions of perfect (but incorrect) knowledge \citep{Barry2016, Ewall-Wice2017, Chokshi2024}.

\newcommand{\BS}{\boldsymbol{\Sigma}}

The Bayesian model specification is complete when we suppose a prior on $\mathbfss{b}$. Generally our prior knowledge about the beam is informed by electromagnetic simulation, rarely with reference to an ensemble of possible variations. In the case that a realistic ensemble of variations can be calculated or measured, their statistics can be encapsulated in this prior. For example, simulating beams with perturbations to feed position and orientation compared to the ideal scenario \citep{Kim2022, Kim2023}, and mapping them to their coefficients in the linear basis can generate a prior for coefficients $\mathbfss{b}$ so long as one is willing to prescribe a probability measure in the space of perturbations. On the other hand, we find in our simulations that the likelihood function by itself is extremely informative about the beam, and so a very strict treatment of the prior may be totally unnecessary. If we suppose a Gaussian prior,
\begin{equation}
    \mathbfss{b} \sim \mathcal{N}\left(\boldsymbol{\mu}, \boldsymbol{\Sigma}\right),
\end{equation}
then the posterior inference is fully analytic:
\begin{equation}
    \mathbfss{b} | \mathbfss{d}, \mathbfss{N}, \mathbfss{I} \sim \mathcal{N}\left(\boldsymbol{\mu}_\mathrm{post}, \boldsymbol{\Sigma}_\mathrm{post}\right)
\end{equation}
where
\begin{equation}
    \boldsymbol{\Sigma}_\mathrm{post}^{-1} = \mathbfss{F}^\dag \mathbfss{N}^{-1}\mathbfss{F}  + \boldsymbol{\Sigma}^{-1}
\end{equation}
and
\begin{equation}
    \boldsymbol{\mu}_\mathrm{post} = \boldsymbol{\Sigma}_\mathrm{post}\left(\mathbfss{F}^\dag\mathbfss{N}^{-1}\mathbfss{d} + \boldsymbol{\Sigma}^{-1}\boldsymbol{\mu}\right).
    \label{eq:mu_post}
\end{equation}
For this work, we suppose a Gaussian prior centered on a least-squares fit to the power beam, with a diagonal $\BS$ with variances equal to the square of this fit i.e. representing $\sim 100\%$ fractional uncertainty in all modes. We have found inferences with this prior to be visually indistinguishable from inferences with a flat prior. However, this is not generally a representation of commonly reported prior knowledge of instrumental beam patterns. Typically, more confidence is expressed in the main lobe compared to the sidelobes \citep{Line2018, Chokshi2024, Kern2025}. Encoding this into the prior would require more complicated structure in the covariance matrix. However, we find that the information in the likelihood outweighs e.g. percent-level mainlobe uncertainties or $100\%$ fractional uncertainties in the sidelobes,\footnote{That is, the posterior standard deviation in the relevent parts of the beam is smaller than these figures by orders of magnitude, implying most of the information is coming from the likelihood} and therefore we do not expect a more stringent prior to significantly affect the results of these numerical experiments. In practice, beam inferences with lower SNR visibilities than our simulated ones may be more reliant on a prior, and including uncertainty in the sky model will probably affect this as well \citep{Kern2025}.

\newcommand{\bd}{\mathbfss{d}}
\newcommand{\NN}{\mathbfss{N}}
\newcommand{\bomu}{\boldsymbol{\mu}}

To check the quality of this model, we perform a posterior predictive check \citep{Gelman2021} against simulated visibility data. In a posterior predictive check, we compare $P(\bd_\mathrm{pred} | \bd, \NN_\mathrm{pred}, I)$ to observed (here simulated) data. That is to say, we compare predictions of where new data would land, marginalized over what was learned about the beam from $\bd$ and using the same sky model. Since there is remaining ambiguity about where this data would land both from model uncertainty and noise, we represent this prediction as a random variable, $\bd_\mathrm{pred}$, with probability distribution $P(\bd_\mathrm{pred} | \bd, \NN_\mathrm{pred}, I)$, where $\NN_\mathrm{pred}$ captures the noise covariance in the experiment to which these predictions are compared. This type of check is more convincing with held out data (data that was not used for the inference), since by its nature the inferred beam was fit to \bd. We observe that tight beam constraints are possible with relatively little data, and that reserving a hold out set should be practically feasible. Under the assumptions listed above the posterior predictive distribution (PPD) is given by
\begin{equation}
    \bd_\mathrm{pred} | \bd, \NN_\mathrm{pred}, I \sim \mathcal{N}\left(\FF_\mathrm{pred}\bomu_\mathrm{post}, \NN_\mathrm{pred} + \FF_\mathrm{pred}\BS_\mathrm{post}\FF_\mathrm{pred}^\dag\right),
    \label{eq:ppd}
\end{equation}
where $\FF_\mathrm{pred}$ indicates a design matrix that is potentially different from $\FF$ only to make accommodations for predicting data with different observational settings but that otherwise obey the same assumptions. For example, one may hold out data at different sidereal times than were used in the beam inference, and compare the PPD to data at those held out times (under the assumption that the beam was indeed the same between the two data sets). 

\section{Simulation Properties}
\label{sec:sims}

\begin{figure*}
    \centering
    \includegraphics[width=\linewidth]{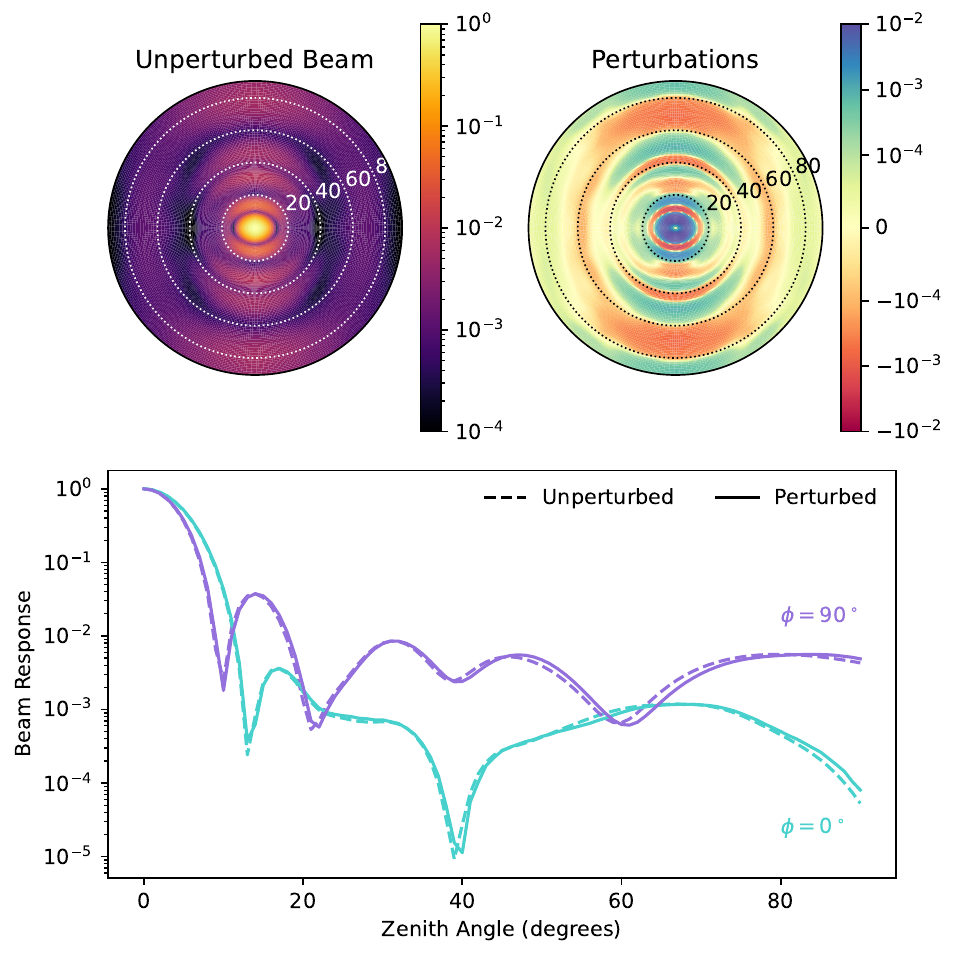}
    \caption{Top: Projection of unperturbed HERA Phase II power beam (left) and the perturbations to this beam that were used in the visibility simulation (right). Absolute perturbations are largest in the main lobe due to the coordinate stretching, while fractional perturbations are largest in the sidelobes (a few percent). Bottom: Slice through the unperturbed beam (dashed) at azimuths of $0^\circ$ (turquoise; horizontal slice in projection plots) and $90^\circ$ (purple; vertical slice), as well as for the perturbed beam (solid). The overall shapes between the two beams are extremely similar by design. The coordinate stretching moves the nulls in the perturbed beam outward in zenith angle by roughly a degree or less.}
    \label{fig:input_beam}
\end{figure*}

We simulate data from a HERA-like array. We use 37 antennas in a close-packed regular hexagon with spacing 14.6 meters. This is a hexagon whose outer layer contains 4 antennas per side and whose longest diagonal contains 7 antennas. Each antenna has an identical power beam, which is an arbitrarily perturbed version of the East-West polarized arm of the HERA Phase II Vivaldi feed (Figure \ref{fig:input_beam}).

We generate the perturbed beam as follows, in a way very similar to beam perturbations that were explored by \citet{Choudhuri+21}. Choices of constants (e.g. offsets in the arguments of functions) are made empirically to generate suitable perturbations. First, we fit the unperturbed beam to an extremely large number of basis functions (7280), including all radial (Bessel) modes out to $k=80$ and all azimuthal modes out to $m=\pm45$. This fits the beam to extreme precision (see Paper I for an example of the precision when fitting the Jones elements with this number of basis functions). We denote these fitted coefficients as $a_{km}$. We then stretch the coordinate system elliptically by defining
\begin{equation}
    \rho'(\theta, \phi) = \rho(\theta)\sqrt{\left(\frac{\cos(m\phi)}{a}\right)^2 + \left(\frac{\sin(m\phi)}{b}\right)^2}.
\end{equation}
for $a=1.01$ and $b=1.02$, and when evaluating the perturbed beam at a given zenith angle and azimuth, we use
\begin{equation}
    B_\mathrm{pert}(\theta, \phi) = g(\rho'(\theta, \phi))\sum_{k=1}^{80}\sum_{m=-45}^{45}a_{km}\frac{J_0\left(u_{k}\rho'(\theta, \phi)\right)e^{im\phi}}{q_{k}}
    \label{eq:Bpert}
\end{equation}
where $g(\rho')$ is a special sidelobe modulation function, defined by the following relations:
\begin{equation}
    g(\rho') = 1 + \sigma_\mathrm{SL}c_\mathrm{SL} h(\rho') K(\rho')
\end{equation}
\begin{equation}
    K(\rho') = \sum_{n=1}^8c_n\sin\left(2\pi \alpha n\rho'\right)
\end{equation}
\begin{equation}
    c_n \sim \mathcal{N}\left(0, 1\right)
\end{equation}
\begin{equation}
    c_\mathrm{SL} = \left(\max\left(K(\rho')\right) - \min\left(K(\rho')\right)\right)^{-1}
\end{equation}
\begin{equation}
    h(\rho') = \frac{1}{2}\left[1 + \tanh\left(\frac{\theta'(\rho') - \frac{\pi}{10}}{\pi/60}\right)\right]
\end{equation}
\begin{equation}
    \theta'(\rho') = \cos^{-1}\left(1 - (\alpha\rho')^2\right).
\end{equation}
The purpose of $c_\mathrm{SL}$ is to make sure that the summation over sinusoids has a controllable dynamic range with the $\sigma_\mathrm{SL}$ parameter. We show the case of $\sigma_\mathrm{SL}=0.2$. The $\tanh$-based $h(\rho')$ ensures that the modulation does not affect the main lobe, which we primarily wanted to be affected by the coordinate stretching. While intricately generated, these perturbations are slight, generating per cent level perturbations in the main lobe and a few per cent level perturbations in the sidelobes. Successful recovery of the perturbations then represents a sensitive test of the method. 

We simulate 10,000 point sources uniformly on the celestial sphere, with flux densities drawn from a log-flat distribution ranging from $10^{-1}$ to $10^2$ Jy at 100 MHz (i.e. a power law distribution with index $-1$). We scale these to 150 MHz with a spectral index of -2.7, which gives a log-flat distribution of fluxes from roughly 33 mJy to 33 Jy at 150 MHz. We then simulate visibilities for 6 hours of sidereal time, sampled at a cadence of 6 minutes. We add noise to the visibilities equivalent to 1000 seconds of coherent integration and 100 kHz bandwidth (e.g. 100 nights of stacked data where each night had a 10 second visibility cadence, similar to HERA). That is, we set the noise covariance for each time, $t$, and antenna pair, $jk$, by
\begin{equation}
    \sigma^2_{tjk} = \frac{V_{jj}(t)V_{kk}(t)}{\Delta T \Delta \nu}
\end{equation}
where $V_{jj}$ is the simulated autocorrelation, $\Delta T$ is the integration depth, and $\Delta\nu$ is the channel width. We then draw noise samples consistent with this covariance and add it to the visibilities. We then consider the noise covariance as known, and do not use the autocorrelations for inference.

We use half these noisy visibilities (the even-indexed ones) for beam inference, and the other half for a posterior predictive check. This means six hours of sidereal time at a cadence of 12 minutes contributes to the beam inference. We found this strategy to produce more accurate reconstructed beams compared to using the same number of time samples spread over a shorter time interval (e.g. 1 hour at a 2-minute cadence). Physically, we find this to be intuitive, since conditioning on the sky model essentially turns all the sources into calibrator sources, and longer ranges of sidereal time let each source trace out longer tracks through the beam. The inference is translated through Fourier-Bessel space, so the analogy is somewhat strained, but may offer an explanation regardless. 

\section{Results}
\label{sec:results}

In this section we demonstrate analysis of beam inference from our simulated data. We start the analysis in Fourier-Bessel space, move into the image domain, and then show comparisons of visibility prediction using the inferred and unperturbed beam models. Lastly, we examine the effects of incomplete sky models on the inference. 

\subsection{Fourier-Bessel posterior analysis}

\newcommand{\bs}{\boldsymbol{\sigma}}

We begin by showing the update from prior to posterior in Fourier-Bessel space. The basic posterior inference consists of calculating $\boldsymbol{\mu}_\mathrm{post}$ and $\boldsymbol{\Sigma}_\mathrm{post}$ once the simulated data have been read into memory. With the statistical model of \S\ref{sec:math} and data volume mentioned in \S\ref{sec:sims}, we observed run times of one to two minutes with 6 Intel E5-2683 v4 Broadwell CPU cores and less than 24GB of RAM. The compute time is dominated by calculation of \mathbfss{F},\footnote{We consider this important to include in the inference time since this would need to be recomputed for each sample in the Gibbs sampling framework described in Appendix \ref{app:Gibbs}.} which naively scales quadratically in number of antennas and linearly in the number of sky sources, times, and beam parameters. 

\begin{figure}
    \centering
    \includegraphics[width=\linewidth]{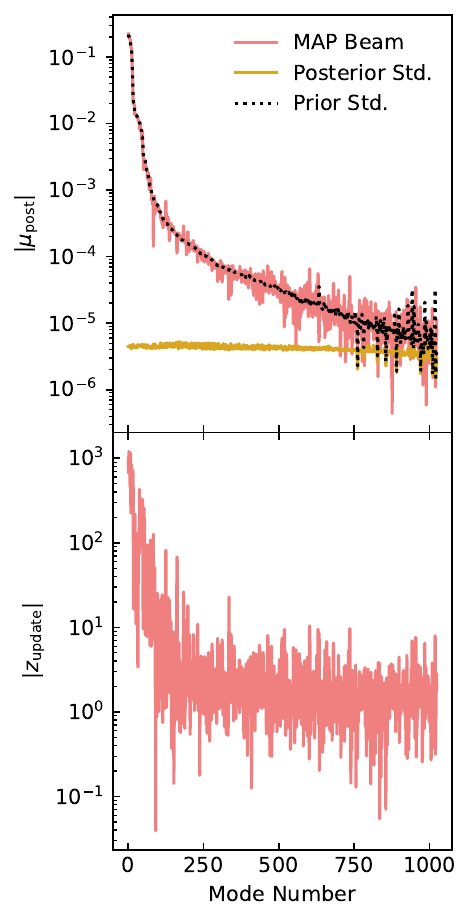}
    \caption{Top: Amplitude of the maximum \textit{a posteriori} (MAP) Fourier-Bessel coefficients (red), along with the posterior standard deviation for each mode (gold), and the prior standard deviation for each mode (black dotted), which are equal to the amplitudes of the prior mean. The vast reduction in uncertainty for the majority of modes indicates strong constraints from the data. Bottom: Amplitudes of $\mathbfss{z}_\mathrm{update}$, which measures the statistical significance of the deviations between prior mean and MAP solution.}
    \label{fig:FB_coeff}
\end{figure}

The top panel of Figure \ref{fig:FB_coeff} shows the amplitudes of the maximum \textit{a posteriori} (MAP) coefficients, along with the posterior standard deviation for each coefficient, as well as the prior standard deviation, which we set equal to the amplitudes of the prior mean. Due to the extreme smallness of the perturbations, the prior and posterior mean are nearly visually identical on this plot (though it does not capture variation in the complex phases of the coefficients). On the other hand, the posterior standard deviations are fairly constant and roughly $4.5\cdot10^{-6}$. The vast reduction in variance from prior to posterior for the first few hundred modes implies that the data has very large constraining power for those basis functions. Until the very highest mode numbers, the ratio of prior to posterior variance is greater than 2 with a few exceptions, indicating stronger contributions from the likelihood than the prior. 

We use the lower panel to plot a measure of the statistical significance of the update from prior to posterior,
\begin{equation}
    \mathbfss{z}_\mathrm{update} = \frac{\left(\bomu_\mathrm{post} - \bomu\right)}{\bs_\mathrm{post}}.
    \label{eq:z_update}
\end{equation}
and
\begin{equation}
    \bs_\mathrm{post} = \sqrt{\mathrm{diag}\left(\BS_\mathrm{post}\right)}.
\end{equation}
In the above two equations, the division and square root operations are done elementwise. In other words, $\mathbfss{z}_\mathrm{update}$ is a vector of $z$-scores of the deviation between the prior and posterior mean, measured in posterior standard deviations of the relevant FB mode. We find that the first hundred or so modes have highly significant updates in excess of $10\sigma$. The majority of the remaining updates are of mild significance or less, with the occasional outlier. It is not necessary or expected for the posterior to update every single basis function. The simulated perturbations are very smooth spatially, and probably have their dominant contributions from the smoother basis functions. The smoothness of the basis functions in this order, which was sorted by the contribution to a least-squares fit of the unperturbed beam, is highly complex. In other words, the number of (radial or azimuthal) oscillations in the basis functions varies considerably (but not totally randomly) from mode-to-mode when examined in order of increasing fit contribution. However, there is a general trend of decreasing smoothness as a function of mode number, and it stands to reason that the unperturbed and perturbed beams might simply agree on these higher-order modes to within the posterior uncertainty. 

\begin{figure*}
    \centering
    \includegraphics[width=\linewidth]{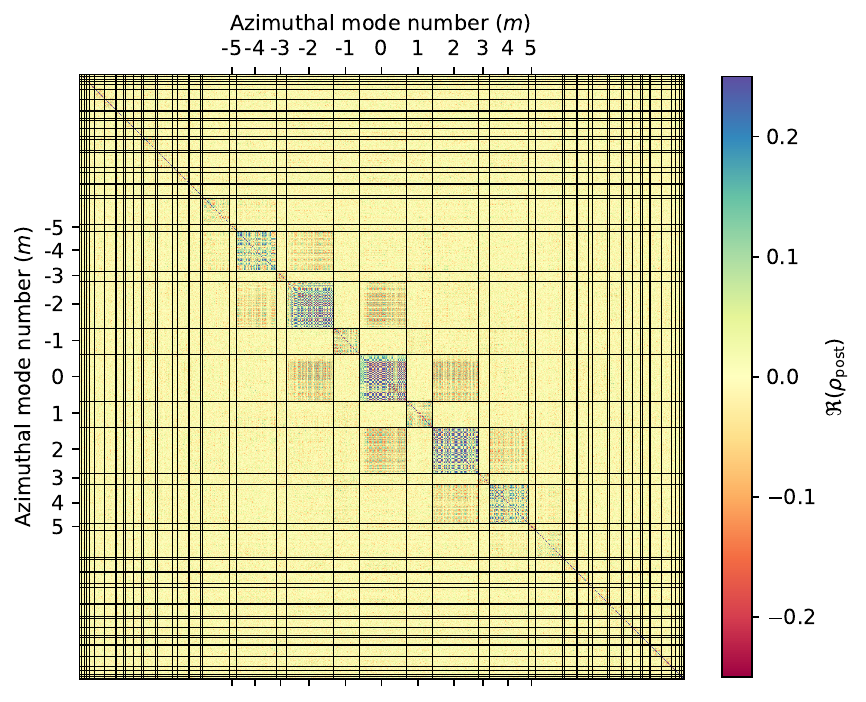}
    \caption{Real part of the correlation matrix for the 1024 complex model parameters, ordered by azimuthal mode numbers (only a subset of the mode numbers are labeled for cleanliness of presentation). Horizontal and vertical black lines indicate boundaries where the azimuthal mode number increments. Most correlations are small, and are strongest among modes that have a common azimuthal mode number. }
    \label{fig:real_corr}
\end{figure*}

It is conventional to illustrate the shape of the posterior distribution of model parameters using a corner plot based on samples. This is visually impractical for a model of this size. Since the posterior is Gaussian, we instead summarize the model degeneracies using a plot of the correlation matrix,
\begin{equation}
    \boldsymbol{\rho}_\mathrm{post} = \frac{\BS_\mathrm{post}}{\bs_\mathrm{post}\bs_\mathrm{post}^\mathsf{T}},
\end{equation}
where the division is done elementwise (Figure \ref{fig:real_corr}). What we find is that, generally correlations are weak, with 96.5\% of pairs having correlation coefficients less than 0.1. To illustrate a pattern, we have ordered the modes according to their azimuthal mode number, $m$, which enumerates the number of periods in the azimuthal sinusoid of the given FB mode. For any given $m$, there are often many associated radial modes with different radial mode number $k$, which counts the number of zeros between zenith and $\rho=1$. Choosing this ordering shows that most of the model degeneracy exists between modes that share an azimuthal mode number but have different radial mode number. There is some notable correlation structure between modes with \textit{even} $m$ that are different in azimuthal number by 2, e.g. $m=-4$ and $m=-2$, but modes with odd $m$ seem to lack this structure. The strongest correlation coefficients (other than the diagonals, which are manifestly equal to 1), are roughly equal to 0.4. We have chosen a slightly more compressed scale than this maximum in Figure \ref{fig:real_corr} to show more structure. Interestingly, some of the modes in the $m=(0, \pm2)$ subblocks are lacking significant correlation, whereas many of the modes in the subblock are correlated. 

\begin{figure*}
    \centering
    \includegraphics[width=\linewidth]{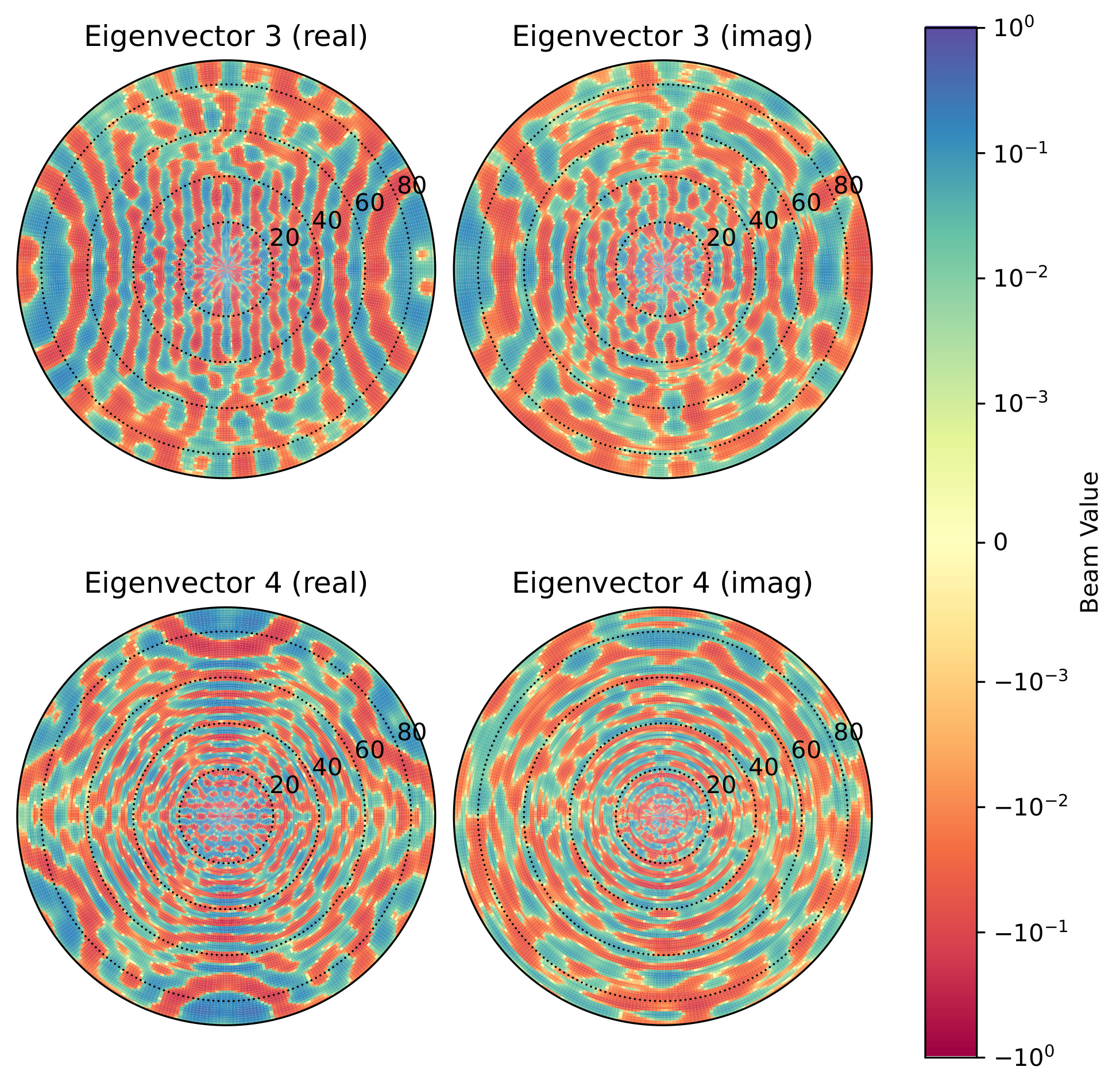}
    \caption{Real and imaginary components of some representative eigenvectors of the posterior covariance with small eigenvalue (i.e. spatial components of the beam pattern that are well-constrained by the data), projected to sky coordinates. The top row approximately resembles the fringe pattern of the shortest East-West baseline in the array, while the real component of the bottom row approximately resembles the expected point-spread function for the close-packed hexagonal array that we simulated. Since these are functions on the sky that the array naturally measures, it makes sense that these components of the beam are some of the ones with tightest constraints.}
    \label{fig:eigenbeam}
\end{figure*}

We can diagonalize the covariance in order to characterize independent posterior uncertainties. This breaks the covariance into orthogonal directions (the eigenvectors) identifying the principle axes of the probability ellipse in Fourier-Bessel space for the Gaussian posterior. The eigenvalue quantifies the posterior variance along that direction. The eigenmode with largest eigenvector is the one with largest posterior uncertainty (the one with weakest constraint), and vice-versa for the eigenvector with smallest eigenvalue. When we examine the first few eigenvectors with smallest eigenvalues (the best constrained modes), we find that when projected onto the sky they approximately resemble shapes derived from properties of the array, such as the fringe pattern of the shortest East-West baseline (Figure \ref{fig:eigenbeam}, top row), or what looks to be the array post spread function (Figure \ref{fig:eigenbeam}, bottom left). We find this physically intuitive in the sense that the most redundant baseline appears as the direction of strongest constraint,\footnote{All three basic 14.6m  baseline orientations appear with nearly equal eigenvalue.} and these are functions on the sky that the array naturally measures independently. 

Eigenvectors beyond the first handful can have complicated shapes and are hard to understand intuitively. In particular, the dominant eigenmode appears to be a smooth random field on the sky (not shown). Its shape may be related to the particular pattern of point sources that were simulated, since 10000 sources on a celestial sphere is a relatively sparse distribution. While it would be ideal to have a physical parametrization of the beam, and perhaps also desirable to have basis functions that the visibilities most naturally constrain, excluding the linear combinations of FB modes represented by the dominant eigenmodes of the covariance matrix would lead to significant errors in this mock analysis. This presents a potential complication. Lacking a straightforward physical parametrization of the beam, it may be prudent to use certain basis functions even if they have no obvious physical correspondence. However, this could lead to overfitting in practical settings when there are sources on the sky not included in the model. This, along with the effects of sky model incompleteness explored in \S\ref{sec:incomplete_sky} suggest joint beam and sky modeling as an important avenue of future research.

The ratio of the largest eigenvalue to the smallest eigenvalue is the condition number, which is a measure of how ill-conditioned i.e. degenerate the covariance matrix is. The condition number of $\BS_\mathrm{post}$ in this mock experiment is roughly 3700, meaning the matrix is not particularly ill-conditioned. While this condition number implies some precision loss in linear operations, this reflects only mild degeneracies among the parameters (roughly speaking a matrix with condition number of $10^{16}$ is essentially singular in double precision). This is important for sampling purposes, where a linear system involving the posterior covariance must be solved in order to generate samples with the appropriate statistics. We expect this condition number (as well as the eigenmodes of the matrix) to depend on the particular sky model in question, which in a Gibbs sampling scheme between a sky and beam model, may change between iterations of the Markov chain. Poorly conditioned covariance matrices indicate strong degeneracies, which may result in poor mixing for a Gibbs sampler. The observed condition number in this work indicates that this may not be problematic, although more realistic sky models with associated uncertainty may change these results.

\subsection{Inferences in sky coordinates}
\label{sec:sky_domain}

\newcommand{\DD}{\mathbfss{D}}

\begin{figure*}
    \centering
    \includegraphics[width=\linewidth]{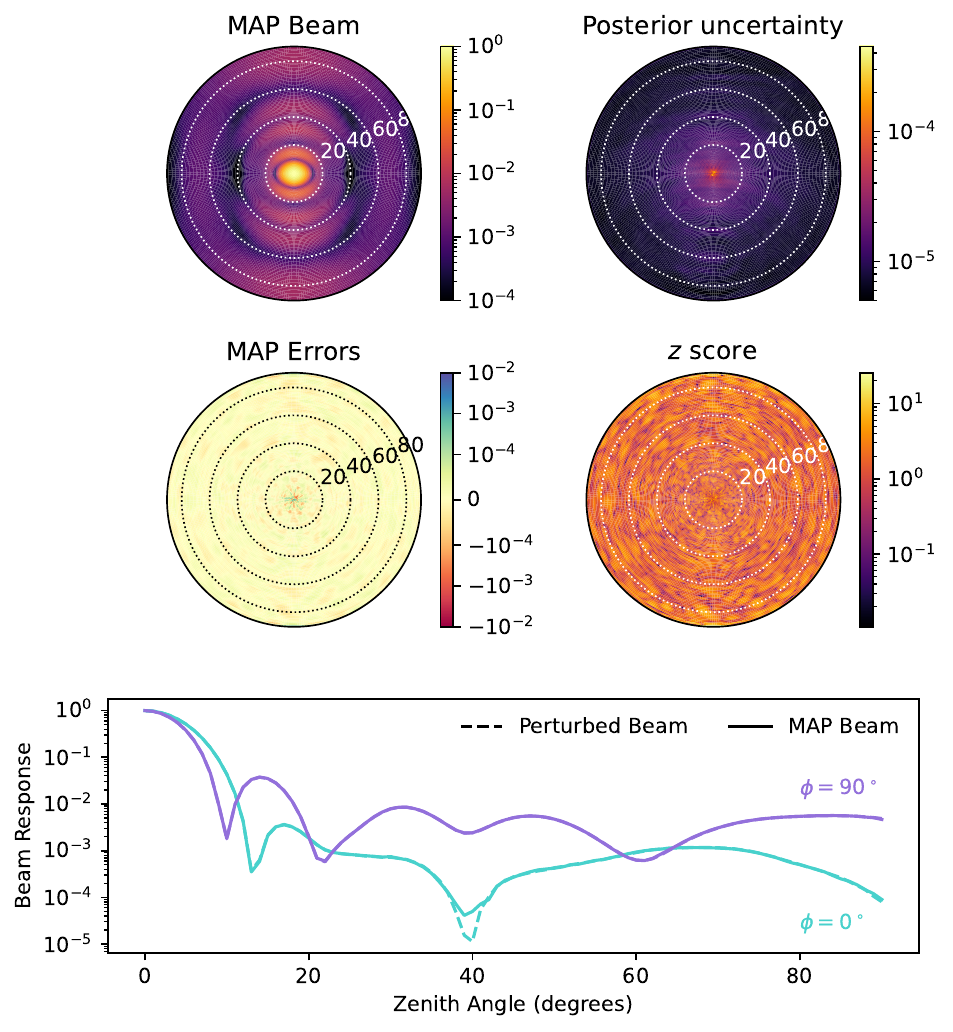}
    \caption{Top left: maximum \textit{a posteriori} (MAP) inferred beam from the visibility simulations. Top right: standard deviation of the posterior as a function of spatial coordinate. They are much smaller than the prior uncertainty, indicating a large amount of information comes from the likelihood. Center left: difference between the input perturbed beam and the MAP beam. Generally errors are much smaller than the size of perturbations (cf. top right of Figure \ref{fig:input_beam}, on the same color scale). Center right: absolute $z$-secore between the MAP beam and the input perturbed beam. Due to model incompleteness, significant statistical tension exists between the inferred beam and input beam at some spatial coordinates (particularly at the horizon), but generally these errors are well-described by the posterior uncertainty (cf. Figure \ref{fig:image_z_score}). Bottom: Slices through the input perturbed beam (dashed) and MAP beam (solid) along azimuths of $0^\circ$ (turquoise) and $90^\circ$ (purple) at 150 MHz. The beams are visibly matching at nearly all positions except the deep null in the $\phi=0^\circ$ slice at $\theta\approx40^\circ$. Compare to bottom panel of Figure \ref{fig:input_beam}.}
    \label{fig:inferred_beam}
\end{figure*}

Next we show the MAP beam in a projection on the sky, along with posterior uncertainty, errors, and absolute $z$-scores of these errors relative to the posterior standard deviations in Figure \ref{fig:inferred_beam}. To make these projections, we define a design matrix $\DD$ that maps from Fourier-Bessel space to a degree-scale grid in $(\theta, \phi)$, where rows of this matrix correspond to different sky positions and columns are the basis functions evaluated at those positions. In other words, $\DD$ implements Equation \ref{eq:beam_expansion} at a a grid of positions on the sky. The MAP solution in this projection is then just
\begin{equation}
    \hat\bb_\mathrm{MAP} = \DD\bomu_\mathrm{post},
\end{equation}
The posterior standard deviation in these coordinates is then
\begin{equation}
    \hat{\boldsymbol{\sigma}} = \sqrt{\mathrm{diag}\left(\DD\BS_\mathrm{post}\DD^\dag\right)}.
\end{equation}
The MAP errors are defined by
\begin{equation}
    \hat{\boldsymbol{\varepsilon}}_\mathrm{MAP} \equiv \hat\bb_\mathrm{input} -\hat\bb_\mathrm{MAP},
\end{equation}
where $\hat\bb_\mathrm{input}$ is evaluated by Equation \ref{eq:Bpert}. Finally, the absolute $z$-scores are
\begin{equation}
    \mathbfss{z} = \frac{\hat{\boldsymbol{\varepsilon}}_\mathrm{MAP}}{\hat{\boldsymbol{\sigma}}}
\end{equation}
where this division is done elementwise.

We see that the inferred beam very closely resembles the input perturbed beam, with errors generally smaller than $10^{-4}$ of the peak beam value. Note that we have made no special effort to ensure that the peak value is equal to 1, or to ensure that the beam is real valued. The inferred beam actually has a tiny imaginary component ($\lesssim 10^{-5}$ everywhere), and we are just showing the real component. One can actually enforce that the beam is real-valued by demanding that azimuthal modes with mode number $m$ always be accompanied by another basis function with azimuthal mode number $-m$, and then mandating that
\begin{equation}
    a_{k,m} = a_{k,-m}^*.
\end{equation}
This condition be accomplished approximately by splitting the inference problem into real and imaginary components and putting the appropriate correlation structure into the prior covariance, $\BS$. It can also be accomplished \textit{exactly} by using cosine and sine modes instead of complex exponentials for the azimuthal modes, and constructing a design matrix such that the cosine modes (and $m=0$ mode) receive no contribution from the imaginary component of $\bb$ and the sine modes receive no contribution from the real component of $\bb$. A similar strategy was used by \citet{Glasscock} to ensure that inferred diffuse maps were real valued despite being expressed in terms of complex spherical harmonics. This necessarily increases the compute involved in the inference, but the explicit constraint is likely to result in more accurate inferences where it is physically motivated. We elect not to do it here because we think it would have very little effect on our conclusions, which are mostly qualitative, and because the beam pattern being real-valued only arises as a consequence of the somewhat restrictive assumption of identical antennas and consideration of the co-polarized visibilities only ($p=q$ in Equation \ref{eq:lin_exp_vis}). However, we consider the interplay between per-antenna beam patterns and this assumption as well as full inference of per-antenna Jones elements to be important topics of future work. 

The posterior standard deviation in sky coordinates is very small -- less than $10^{-4}$ in most places. This is mainly due to the fact that the simulated visibilities have very low noise, but is also partially determined by the fact that the system of coefficients is overconstrained (i.e. there are fewer free parameters than measurements). We suspect that the uniformity of this small uncertainty is mainly owed to the fact that the sky model is complete and also statistically isotropic. Importantly, we found that beam uncertainties were extremely large in preliminary experiments where we did not simulate sources all the way to the horizon and used a flat prior on the beam coefficients. Note that this is different from simulating sources down to the horizon, but only including a subset of them in the sky model for inference. Sky model incompleteness is known to impart spectral structure in direction-independent calibration of radio interferometers \citep{Barry2016}, and it is likely that beam inference would be strongly affected by an incomplete or inaccurate sky model. We perform a cursory examination of that in \S\ref{sec:incomplete_sky}.

\begin{figure}
    \centering
    \includegraphics[width=\linewidth]{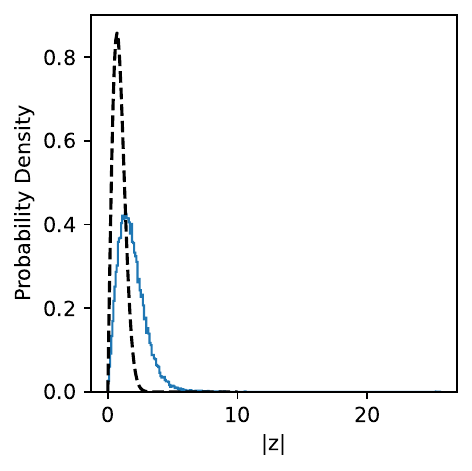}
    \caption{Histogram of absolute $z$-scores corresponding to bottom right panel of Figure \ref{fig:inferred_beam}, along with a Rayleigh distribution corresponding to expectations of an inference using a complete beam model (dashed). }
    \label{fig:image_z_score}
\end{figure}

The perturbed beam cannot be exactly represented on this grid with the basis functions that we chose. That is, a least-squares fit to the chosen basis functions on this grid produces nonzero residuals. This means that the inferred beam cannot exactly represent the perturbed beam either; there are bound to be errors even in the limit of zero noise. Said another way, the model is explicitly assuming that the beam \textit{can} be exactly represented, and since this assumption is not obeyed, the statistical uncertainty as represented by the posterior distribution does not exactly characterize the observed distribution of errors. We can see this in the bottom-right panel of Figure \ref{fig:inferred_beam}, where there are a handful of regions with large $z$-scores with respect to the posterior standard deviation, e.g. in excess of $20\sigma$. This indicates that there is indeed statistical tension between the input beam and the inferred beam. However, the overall distribution of absolute $z$-scores in this space is not wildly different from the expected Rayleigh distribution, as shown in Figure \ref{fig:image_z_score}. In summary, the model, including the statistical uncertainty reflected by the posterior distribution, accommodates the perturbed beam well with the exception of some outliers.

Slices through the input and inferred beams at two azimuths are shown in the bottom panel of Figure \ref{fig:inferred_beam}. There is strong agreement between the inferred and input beams. Errors are visibly much smaller than the perturbations themselves (cf. Figure \ref{fig:input_beam}). The most visible difference is a misestimated null, however the largest absolute errors happen within the main lobe, as evidenced by the center left panel of Figure \ref{fig:inferred_beam}. 

\subsection{Visibility prediction check}

\begin{figure*}
    \centering
    \includegraphics[width=\linewidth]{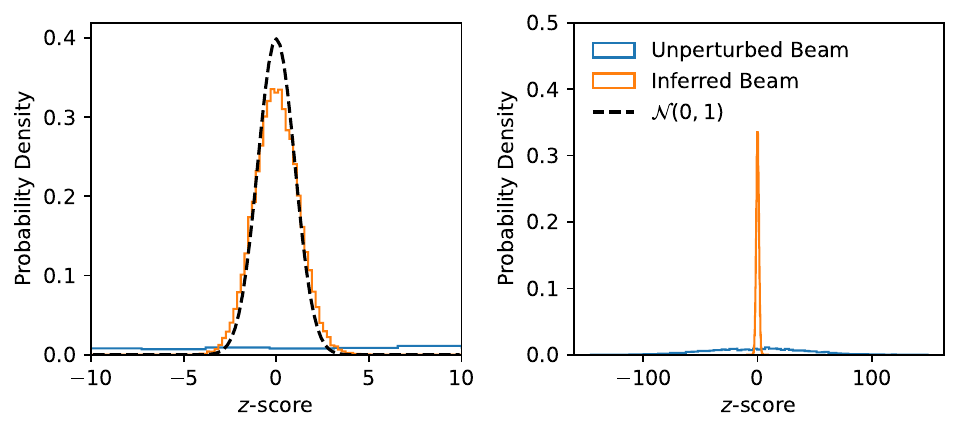}
    \caption{Residuals between the predicted and simulated (real and imaginary components of the) visibilities at times not used for inference, using the inferred beam normalized by the standard deviation of the posterior predictive distribution (PPD; orange), or using the unperturbed beam and just noise standard deviation (blue). The variance of the PPD is dominated by the noise variance term and therefore does not account for the large statistical discrepancy between these two models. Predicted visibilities using the inferred beam model are generally statistically consistent with the simulated ones, whereas using the unperturbed beam to make predictions results in extremely large outliers relative to the simulated visibilities. Both panels contain identical histograms and are just shown for better feature identification.}
    \label{fig:vis_res_hist}
\end{figure*}

To test the fidelity of the inferred beam in a data-driven way, we perform a posterior predictive check. We demonstrate this type of check in particular because it is a way statistical inferences can be validated in practice without absolute knowledge of the true beam. In other words, we are demonstrating a model comparison and validation technique that can be performed in observational practice, with beams inferred from measured visibilities. It involves a comparison of the statistical expectations of visibilities held out from inference,\footnote{In general, posterior predictive checks do not have to be performed on held out data, but models are biased to make better predictions for data they were fit to, so doing it in this way eliminates potential worries associated with that.} based on what was learned from the data used for inference. The basic idea is that if the \textit{a posteriori} model is accurate and associated uncertainties are well-characterized, the held out data will perform well according to various statistical tests (e.g. $\chi^2$) with the posterior predictive distribution (PPD; Equation \ref{eq:ppd}) used for the null hypothesis. 

To do this check, we first form the PPD for the visibilities at sidereal times not used for the beam inference. We then compare to the residual between the mean of the PPD and the observed visibilities, normalized with respect to the standard deviation of the PPD for each complex component (we form posterior predictive $z$-scores). Note that the noise variance dominates the variance of the PPD since the beam is so well-constrained by the initial set of visibilities. For a perfect model, these $z$-scores would be distributed according to a standard normal distribution. We observe in Figure \ref{fig:vis_res_hist} that this is almost true, suggesting that there is a barely detectable difference between the predicted and simulated visibilities. For comparison, we also perform the same exercise using the unperturbed beam, by simulating visibilities using the unperturbed beam, taking the residuals with the original simulated visibilities, and normalizing with respect to just the noise standard deviation. We see that the $z$-scores using the unperturbed beams are extremely large, ranging up to $100$ in some instances. In other words, a prediction of the visibilities using the unperturbed beam demonstrates strong statistical discrepancies with the original simulated, visibilities at this integration depth, despite the fact that the absolute perturbations in the beam are small. 

\subsection{Incomplete sky model}
\label{sec:incomplete_sky}

A generic feature of fitting models with free parameters is that the free parameters may compensate for inaccuracies in features of the model that are fixed, and in potentially unpredictable ways. Since it is rarely reasonable to consider a model as a perfect description of the data, it is important to investigate the impact of model imperfections. One assumption we made (and obeyed) in previous sections is perfect knowledge of sources on the sky. In particular, we simulated a catalog of point sources with perfectly known fluxes and positions. There was no diffuse component, and we had no missing sources. These assumptions do not perfectly represent the state of knowledge of radio astronomy, where point source catalogs are only complete down to some nominal flux value, and generally only survey a limited view of the sky \citep[e.g.][]{LoTSS, GLEAMX2}. Furthermore, diffuse emission is a particularly significant component at longer wavelengths such as in observations focused on finding the cosmic reionization signal \citep{Byrne2022b, BGSM, BGSM2}. In a Bayesian setting, we handle sky model inaccuracies by ascribing uncertainties to various aspects of existing models and constrain them further with the data where possible. That far exceeds the scope of this work, where our primary goal is to explore beam inference using linear basis expansions at a basic level. Instead, we will briefly explore what happens to the inferred beam when a small amount of flux is missing from the assumed sky model, in order to motivate fully Bayesian joint modeling the sky and beam in future ventures. 

To explore this, we use the same simulations and beam prior as before, but change the form of the FB-to-visibility design matrix in the forward model that relates the sky and beam to the visibilities (Equation \ref{eq:forward_model}). Specifically, we set all source fluxes to zero that are in the bottom order of magnitude in the flux distribution. Since the flux distribution is log-uniform, this removes approximately 1\% of the total flux from the catalog. This emulates the effect of having missing sources in the flux catalog below some limit, and proceeding as if one had complete knowledge of the sky anyway. We examined all of the same plots as in the previous subsections, but only display and discuss a subset of them in this one.

\begin{figure*}
    \centering
    \includegraphics[width=\linewidth]{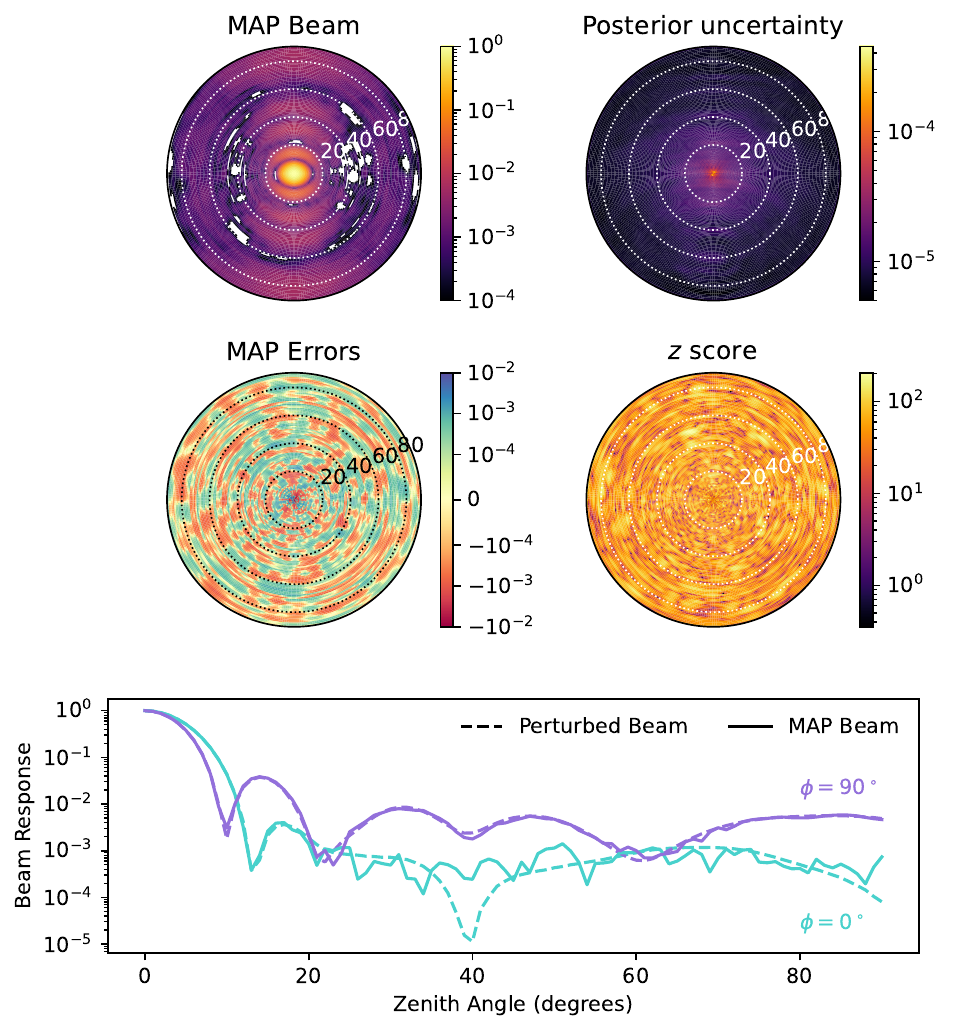}
    \caption{Same as Figure \ref{fig:inferred_beam}, but in the case of using an incomplete sky model to infer the beam, and with the exception that in the slice plot we are showing the beam amplitude instead of its real component. The upper left panel is showing the real component of the inferred beam; white pixels indicate negative values, which we interpret as a compensatory mechanism for the missing sources. Beam inferences are reasonably accurate down to roughly $10^{-2}$, beneath which the inferences become quite noisy. The calculated posterior uncertainty does not resemble the actual distribution of errors. }
    \label{fig:recon_missing_sources}
\end{figure*}

Conditioning on an incomplete source catalog results in practically and statistically significant errors in the inferred beam. We show a multipanel figure of beam inference metrics in sky coordinates in Figure \ref{fig:recon_missing_sources}, in the same style as Figure \ref{fig:inferred_beam} but with one difference: the bottom panel shows the slices through the beam amplitude rather than just the real component. In this case, the imaginary component of the MAP beam is substantially larger ($\lesssim 10^{-3}$ everywhere). While the real component of the beam largely resembles the perturbed beam, sidelobes below $10^{-2}$ appear noisy, and there are regions of negative real component in the deeper sidelobes (indicated by the white pixels in the upper left panel of Figure \ref{fig:recon_missing_sources}). The structure of the main lobe and first sidelobes at different azimuths appear largely preserved. Furthermore, the slices through the beam amplitude show that the amplitude of the inferred beam is roughly correct in most places. The posterior uncertainty is nearly identical as the perfect sky model case, but due to the significantly larger errors, does not adequately characterize their statistics.


In this numerical experiment, the beam model compensated for the missing sources by applying direction-dependent phases, roughly preserving amplitude. Using the MAP beam wholesale may then be particularly disastrous in the search for faint cosmic signals, where the spatial structure of the target signal, which is not included in the sky model essentially by definition, may be scrambled by this rephasing. Having only performed this at one frequency, we do not know how this might also couple to the spectral axis, along which it is critical to keep a high dynamic range for 21cm cosmology \citep{Morales2012, Liu2014, Barry2016, Trott2016}. Of course, both of these problems in part stem from the fact that we have made no significant effort to ensure that the beam is real and positive via the prior, which are basic properties of power beams of arrays with identical receiving elements. In other words, this is a bit of a ``slippery slope'' argument in the sense that there is unlikely to be an analysis team that would take this MAP solution through the rest of a pipeline without attempting to correct these features. Instead, this paper serves to show that while a sufficiently accurate sky model may alleviate this issue, or some Bayesian miracle may occur when marginalizing over sky model uncertainty, it may be more prudent (or even critical) in future endeavors to solve them via careful specification of the beam model.

As mentioned in \S\ref{sec:sky_domain}, enforcing real-valuedness of the beam in the sky domain is easily accomplished through a special construction of the design matrix. Establishing positive-definiteness in a (Gaussian) \textit{linear} model is significantly more difficult if strict linearity is to be maintained. For example, taking the absolute value of a linear combination of basis functions can enforce positive-definiteness in a way that accommodates Bayesian inference, but leads to a less tractable posterior due to the nonlinearity of the absolute value function \citep{Kern2025}. On the other hand, maintaining linearity allows for a wider swath of sampling techniques to be employed that may allow for rapid traversal of ultra-high dimensional parameter spaces. 

Theoretically, one can enforce beam positivity by just truncating the prior in image space, and mapping this truncation back to Fourier-Bessel space. However, the mapping between positive-definiteness in sky coordinates and the relevant region in FB space is not straightforward, and some practical issues arise. While theoretically there should be some high-dimensional polytope in FB space that maps to a subspace of positive-definite beams under the linear transformation reflected by the FB-to-sky design matrix, drawing posterior samples constrained to this polytope is numerically challenging since it may be extremely deep in the tails of the likelihood. For example, some pixels in Figure \ref{fig:recon_missing_sources} are tens to hundreds of posterior standard deviations away from positivity (neglecting the imaginary component for simplicity, which is another issue). This means to draw posterior samples, one would need to draw from a deeply truncated multivariate Gaussian distribution with a nontrivial covariance structure. While sampling strategies exist for such distributions, they can be slow \citep{truncmv1, truncmv2}, and speed is one of the main advantages of a linear model in this context. 

Solving the problems of real-valuedness and positivity of the power beam may result in more realistic beam patterns when conditioning on an inaccurate sky model. However, there is still liable to be some bias in the inferred imparted by the sky model inaccuracies. In theory, Bayesian methods that jointly model an uncertain beam and sky can alleviate some of the pressure of this bias. We have demonstrated that there are model degeneracies between the sky and beam by showing that beam inference is sensitive to the assumed sky model. When there is a degeneracy, the posterior distribution of the beam conditional on a particular sky model can be much narrower than the beam's marginal posterior distribution \citep[e.g.][Figure 9]{Kern2025}. This means in the ideal scenario where it is possible to represent the beam using the chosen model, conditioning on an inaccurate sky model can disallow the beam posterior from encompassing the true beam (this is the bias). Where the beam and sky model are the only inaccurate components of the system, jointly modeling the sky with the beam allows for a broader exploration of the space that mutually updates the sky and beam probability distributions so that they are centered on more accurate values, and so that uncertainty estimates reflect the important relationships between the two model components. We therefore strongly recommend this strategy in future work involving beam inference. 

Joint models of the beam and sky as well as solutions to some of the unitary ambiguities referred to earlier have been explored through the lens of regularized optimization and show excellent promise \citep{Yatawatta2021, Yatawatta2024, Gan2023, Brackenhoff2025}. Regularized optimization often permits a Bayesian interpretation, where the regularization terms act as parameter priors and/or extra hierarchical layers in the model. To extend this to a fully Bayesian approach, we would aim to draw samples from the joint posterior probability distribution, rather than just finding the best (regularized) fit parameters. Such priors often imply a non-Gaussian posterior even in the case of a linear model. This makes fully Bayesian extensions of such work in a basic Gibbs scheme as suggested here somewhat strained, because the conditional beam posterior may no longer be easily sampled. Adaptive rejection sampling steps, e.g. based on the Metropolis algorithm, can be included within a Gibbs sampler while maintaining convergence properties \citep{Gilks1995, Gelman2021}. This presents a potential avenue for sampling from more complicated conditional distributions, however the scaling of such an algorithm in high dimensions may require careful tailoring. On the other hand, \citet{MWGscaling} demonstrate that such schemes may scale better than alternatives such as no-U-turn sampling \citep[NUTS;][]{Hoffman2014} for certain high dimensional models (thousands of parameters) in practical circumstances. Therefore, Gibbs schemes overall remain a highly viable candidate for HDR radio astronomy experiments despite challenges presented in this work.

\section{Summary and Conclusions}
\label{sec:conc}

In this paper, we developed a Bayesian framework for inferring an array-averaged, per-frequency primary beam pattern from interferometric visibilities, given a fixed sky model and a linear basis model for the beam with over 1000 terms. A beam model of this size is sparse relative to a pixel-based basis, but has a high dimensionality compared to typical statistical models. Exploration of such a large model was necessary to ensure that high-fidelity reproduction of realistic, complicated beam patterns is possible. We used 2-dimensional Fourier-Bessel functions as our basis set, but theoretically the approach will work with any linear basis expansion of the beam. Handling sky models with many thousands of sources, and beam models with hundreds to thousands of parameters, is computationally challenging however, due to the need to repeatedly re-simulate the interferometric visibility model -- potentially involving thousands of baselines for close-packed arrays like HERA -- when any of these parameters changes.

Our approach is able to efficiently draw samples from the posterior distribution for the beam parameters (conditioned on the sky model) by making use of a Gaussian constrained realization (GCR) formalism. This permits random draws of samples by solving a large linear system involving a pre-calculated projection operator matrix that goes from beam basis functions to visibilities. This is highly scalable, and we were able to demonstrate efficient and accurate statistical sampling in the case that the sky model is accurate. While errors of order 1\% in the sky model did affect the accuracy of the inference, the recovered beam model was still close to the true model. We also presented a Gibbs sampling scheme (see Appendix \ref{app:Gibbs}) for performing joint sampling of the sky model parameters at the same time as the beam parameters, although a demonstration of this is left to future work.

The significance of this approach is that realistic, detailed beam models with many parameters can be successfully constrained using typical interferometric data taken `in situ', without the need for separate calibration procedures such as drone overflights. There is some reliance on the accuracy of the sky model to make this feasible, but it need not be unrealistically accurate to get reasonable results. Most importantly, the posterior distribution of the parameters is recovered, allowing detailed statistical assessment of the beam model and its uncertainties, and propagation of the uncertainties through the rest of the data analysis, rather than needing to rely on a single `best fit' model with potentially poorly-constrained model errors.

As a specific example, we demonstrated recovery of a perturbed beam based on the HERA Phase II receiving elements in a mock data analysis. Using a toy simulation consisting only of point sources with a log-flat flux distribution, we showed that with sufficient integration depth and a (perfectly) accurate sky model, one can obtain extremely accurate and precise inferences about the beam with limited input from the prior. For our simulations with 37 antennas (resulting in 666 baselines), 60 times, and a single frequency channel, errors were generally less than $10^{-4}$ of the peak beam value, and reasonably well-summarized by the posterior uncertainty. Perturbations that we added to the beam were very slight ($10^{-2}$ in the main lobe, and $\lesssim 10^{-3}$ in the sidelobes), but they were highly statistically significant from a visibility prediction standpoint. That is, inaccuracies that would occur due to using the unperturbed beam were highly resolvable from visibility data, with $z$-scores broadly distributed (as high as 100 in some cases), whereas using the inferred beam demonstrated $z$-scores that were nearly standard normal -- as expected for an accurately inferred model. A more quantitative forecast of the performance in a practical setting would require more realistic modeling of the sky, but we are optimistic that the performance of the method would remain good even with the addition of complications such as diffuse emission into the sky model. 

In terms of statistical properties, posterior inferences were well-behaved. Degeneracies among individual Fourier-Bessel modes are mild, and in this case a Gaussian posterior makes the very large model space relatively easy to summarize and explore. The fully analytic (and Gaussian) nature of this posterior is in part owed to the prior, however since much of the information came from the likelihood, this choice seems like it will be relatively inconsequential. We also found physical correspondence with array characteristics in the eigenspectrum of the posterior covariance, suggesting future avenues for experimental design and model building. In particular, exploration of the posterior covariance in mock analyses such as these may facilitate experimental design for direct beam measurements, such as by informing where visibility-based beam inferences can be supplemented by e.g. drone-based measurements. 

We found that inferences degraded when we removed 1\% of the total flux from the sky catalog used for inference, corresponding to the faintest 1/3 of the simulated sources. This suggests that sky uncertainty should be included in the statistical modeling in order to avoid biased beam inferences. Additionally, applying basic \textit{a priori} constraints such as positivity for array power beams should help produce more realistic beam patterns in spite of sky model inaccuracies, though maintaining this constraint and the fully linear nature of the model may present practical difficulties due to the requirements of characterizing high-dimensional truncated multivariate Gaussian distributions.  An excellent advantage to a fully analytic beam inference solution is that it can be combined with a sky inference framework where conditional posterior samples from the sky and beam are drawn alternately in a Gibbs sampling scheme. Based on this work, the rate-limiting step in such a Gibbs sampler will essentially be repeated visibility simulation under varying sky and beam models. We therefore recommend this as an important supplemental focus to applications of Bayesian inference in radio astronomy.

An important avenue of beam inference that we did not explore is the spectral response (see Paper I). In general, this framework could be run in parallel for multiple frequency channels in the data. However, analyzing the frequency channels jointly allows one to express priors that relate adjacent frequency channels via knowledge of spectral smoothness (or lack thereof). This is especially critical for high dynamic range experiments such as those searching for the cosmic 21-cm signal, where inferring separate frequency channels independently may lead to absorption of the target signal into the inferred beam model. There are various ways of folding this in to the framework presented here, such as by using smooth basis functions for the frequency axis, or enforcing smoothness via the Gaussian prior. This can therefore be explored in future work while maintaining the advantages of our approach.


\section*{Acknowledgements}

We are grateful to Aman Chokshi and Robert Pascua for helpful discussions. This result is part of a project that has received funding from the European Research Council (ERC) under the European Union's Horizon 2020 research and innovation programme (Grant agreement No. 948764).

\section*{Conflicts of Interest}

The authors have no conflicts of interest to declare.

\section*{Data Availability}

The unperturbed HERA beam model is publicly available from \url{https://data.nrao.edu/hera/beams/}. The Hydra sampler software and associated tools are available from \url{https://github.com/HydraRadio/Hydra}.

\bibliographystyle{mnras}
\bibliography{hydra_beam}

\appendix

\section{Gibbs Sampling}
\label{app:Gibbs}

Gibbs sampling is an MCMC method for sampling from multivariate probability distributions. In particular, suppose we want to draw samples from the joint probability distribution of $n$ variables,
\begin{equation}
    P(\mathbf{x}) = P(x_1, x_2, \dots, x_n).
\end{equation}
The most basic form of Gibbs sampling draws samples from this joint probability distribution by drawing samples from the conditional probability distributions of each individual variable in turn. For example, starting from some initial state $(x_1^{(0)}, x_2^{(0)},\dots,x_n^{(0)})$, the first round of samples is drawn with the following recursive pattern
\begin{equation}
    P(x_1^{(1)}) = P(x_1 | x_2=x_2^{(0)}, x_3=x_3^{(0)}, \dots, x_n=x_n^{(0)})
\end{equation}
\begin{equation}
    P(x_2^{(1)}) = P(x_2 | x_1=x_1^{(1)}, x_3=x_3^{(0)}, \dots, x_n=x_n^{(0)})
\end{equation}
\begin{equation}
    P(x_j^{(1)}) = P(x_j | x_1=x_1^{(1)},\dots,x_{j-1}=x_{j-1}^{(1)},x_{j+1}=x_{j+1}^{(0)}, \dots)
\end{equation}
and so on. The set of samples, $\mathbf{x}^{(1)}=(x_1^{(1)}, x_2^{(1)}, \dots,x_n^{(1)})$ define a single joint sample in the Markov chain. Once this joint sample has been collected, the process starts anew, drawing $\mathbf{x}^{(2)}$ using the values of $\mathbf{x}^{(1)}$. Theoretically, this Markov chain satisfies detailed balance, and its samples will be distributed according to the target distribution, $P(\mathbf{x})$, in the infinite limit. In particular, it is a special case of the Metropolis-Hastings method where every sample is accepted \citep{Mackay2003, Gelman2021}.

Assessing convergence of the Markov chain is generally an exercise in heuristics, particularly when large spaces with complicated degeneracies are involved. Empirically, Gibbs samplers of the basic form above can be slow to converge in the face of degeneracies. A simple way to combat this, if $P(\mathbf{x})$ is ammenable, is to perform \textit{blocked} Gibbs sampling, where joint samples of subsets of $\mathbf{x}$ are drawn \textit{simultaneously} from the appropriate conditional distributions. For example, suppose $\mathbf{x}$ can be broken into two subsets $\mathbf{y}_1$ and $\mathbf{y}_2$. If the multivariate conditional distributions $P(\mathbf{y}_1 | \mathbf{y}_2)$ and $P(\mathbf{y}_2 | \mathbf{y}_1)$ are tractable from a sampling standpoint, then forming a Markov chain by alternating samples from these two multivariate distributions will also converge to the target distribution. This works especially well if $\mathbf{y}_1$ and $\mathbf{y}_2$ are relatively independent subsets, even if variables within a subset are highly degenerate. The improvement comes from the fact that these degenerate variables within a subset are drawn jointly.  

Note that Gibbs sampling always relies on the tractability of the lower-dimensional conditional distributions. In some applications, this can be a severe limitation, such as those involving nonlinear models, or simulation-based models. However, Gibbs sampling has enjoyed decades of success through the statistical literature and remains a standard method for Bayesian inference. Examples of Gibbs sampling in the astrophysics and cosmology literature in high-dimensional contexts include the \textsc{Commander} CMB analysis pipeline \citep{Commander}, and recent advancements in analysis of pulsar timing array data \citep{PTAGibbs}. 

\balance

Throughout the main body of this work, we have made reference to folding this primary beam inference framework into a Gibbs sampler where we simultaneously  allow for beam and sky model uncertainties. Namely, we would want to draw samples from the joint posterior probability distribution of sky and beam parameters, $P(\mathbfss{b}_\mathrm{beam}, \mathbfss{b}_\mathrm{sky} | \mathbfss{d})$, where $\mathbfss{b}_\mathrm{beam}$ are parameters in a beam model and similarly for $\mathbfss{b}_\mathrm{sky}$. As an example, consider marginalization over a sky model that is parametrized in terms of spherical harmonic modes, such as in \citet{Glasscock} and \citet{Kern2025}. This type of sky model is particularly synergistic with our beam inference framework from a Gibbs sampling perspective. Examining Equation \ref{eq:power_beam}, we see that there is a duality between the sky and beam models. In particular, if both the sky and beam models are linear, then $P(\mathbfss{b}_\mathrm{beam}|\mathbfss{b}_\mathrm{sky}, \mathbfss{d}, \mathbfss{N})$ and $P(\mathbfss{b}_\mathrm{sky}|\mathbfss{b}_\mathrm{beam}, \mathbfss{d}, \mathbfss{N})$ are multivariate Gaussian distributions (assuming appropriate priors). That is, Equation \ref{eq:likelihood} can take both of the following forms, implying a Gaussian likelihoods for the beam and the sky when the other model components are conditioned on:
\begin{equation}
    \mathbfss{d} | \mathbfss{b}_\mathrm{beam}, \mathbfss{b}_\mathrm{sky}, \mathbfss{N}, \sim \mathcal{N}\left(\mathbfss{F}(\mathbfss{b}_\mathrm{sky})\mathbfss{b}_\mathrm{beam}, \mathbfss{N}\right)
\end{equation}
\begin{equation}
    \mathbfss{d} | \mathbfss{b}_\mathrm{beam}, \mathbfss{b}_\mathrm{sky}, \mathbfss{N}, \sim \mathcal{N}\left(\mathbfss{X}(\mathbfss{b}_\mathrm{beam})\mathbfss{b}_\mathrm{sky}, \mathbfss{N}\right).
\end{equation}
where $\mathbfss{F}(\mathbfss{b}_\mathrm{sky})$ is a rewriting of $\mathbfss{F}(I)$ from Equation \ref{eq:likelihood} in terms of  $\mathbfss{b}_\mathrm{sky}$, and $\mathbfss{X}(\mathbfss{b}_\mathrm{beam})$ is a linear operator acting on the vector $\mathbfss{b}_\mathrm{sky}$ to produce model visibilities that can be determined similarly to the machinations involved in Equation \ref{eq:forward_model}. See \citet{Glasscock} for a working example. This implies that the joint beam and sky modeling problem can be broken into alternating multivariate Gaussian draws, where in each Gibbs step, a large linear system is formed and solved to produce `Gaussian constrained realizations' \citep{Eriksen2007, Kennedy2023}. This is highly advantageous because the full joint distribution of the sky and beam parameters is generically non-Gaussian (since their parameters are multiplied together in the likelihood), but this Gibbs sampling scheme breaks this complicated high-dimensional non-Gaussian sampling problem into a sequence of tractable Gaussian sampling problems. The design matrices \mathbfss{F} and \mathbfss{X} must be calculated anew in each sample. Since the backbone of each of these design matrices is visibility simulation, this emphasizes the need for rapid visibility simulation techniques.

\end{document}